\documentclass[twocolumn,notitlepage,showpacs,
superscriptaddress]{revtex4-1}
\usepackage{graphicx}

\usepackage[colorlinks,citecolor=blue]{hyperref}

\begin{document}
\title{One-way quantum deficit and quantum
coherence in the anisotropic $XY$ chain}

\author{Biao-Liang Ye}
\affiliation{School of Physics
and Electronic Information, Shangrao Normal
University, Shangrao 334001, China}
\author{Bo Li}
\email{libobeijing2008@163.com}
\affiliation{School of Mathematics $\&$
Computer Science, Shangrao Normal
University, Shangrao 334001, China}
\author{Li-Jun Zhao}
\affiliation{School of Mathematical Sciences,
Capital Normal University, Beijing 100048,
China}
\author{Hai-Jun Zhang}
\affiliation{School of Mathematical Sciences,
Capital Normal University, Beijing 100048,
China}
\author{Shao-Ming Fei}
\email{feishm@cnu.edu.cn}
\affiliation{School of Mathematical Sciences,
Capital Normal University, Beijing 100048,
China}
\affiliation{Max-Planck-Institute for
Mathematics in the Sciences, 04103 Leipzig, Germany}

\begin{abstract}
	In this study, we investigate pairwise non-classical
	correlations measured using a one-way
	quantum deficit as well as quantum
	coherence in the $XY$ spin-1/2 chain
	in a transverse magnetic field for
	both zero and finite temperatures.
      	The analytical and numerical results of our investigations are presented.
	In the case when the temperature is zero,
	it is shown that the one-way quantum
	deficit can characterize quantum phase
	transitions as well as quantum
	coherence. We
	find that these measures
	have a clear critical
	 point at $\lambda=1$.
	When	$\lambda\le1$, the one-way quantum deficit
	 has an analytical expression that
	coincides with the relative entropy of
	coherence. We also study an  $XX$ model
	and an Ising chain at the finite
	temperatures.\\

{Keywords: One-way quantum deficit, quantum coherence, quantum phase transitions, $XY$ chain}
\end{abstract}

\pacs{03.67.-a, 73.43.Nq, 75.10.Pq}

\maketitle

\section{Introduction}
Quantum entanglement is considered to be
a resource in many quantum information processing
tasks \cite{Amico2008,Horodecki2009}, and entangled states
have been experimentally created using 14 trapped ions \cite{Monz2011},
five superconducting qubits \cite{Barends2014},
and optical systems \cite{Lin2015a,Heilmann2015,Wang2016}.
Quantum entangled states can been used as resources in
quantum cryptography \cite{Ekert1991},
quantum dense coding \cite{Bennett1992},
quantum communications \cite{Zou2014}, and quantum key
distribution \cite{Cao2015}.
Nonetheless, in the past few decades, it is realized that
quantum correlations beyond entanglement also play
essential roles in
quantum information processing \cite{Modi2012}.
Many measures have been proposed  to quantify
quantum correlations in physical systems, e.g.,
quantum discord \cite{Ollivier2001,Henderson2001},
one-way quantum deficit \cite{Oppenheim2002},
measurement induced disturbance \cite{Luo2008},
geometric discord \cite{Dakic2010},
quantum dissonance \cite{Modi2010}, and
measurement induced non-locality \cite{Luo2011}.

In recent years, quantum coherence
has also attracted considerable attention
\cite{Streltsov2016}; consequently, reasonable measures of quantum coherence
have been discussed extensively \cite{Baumgratz2014}.
 As an analogy of quantum
entanglement, quantum coherence may be also
considered as a resource to characterize the
classical-quantum boundary \cite{Adesso2016}.

The study of various quantum correlation
measures in the ground states of spin models
has been an active area of research.
Entanglement in the
finite size $XY$ chain has been investigated \cite{Osborne2002}.
Multi-particle entanglement in an anisotropic
$XY$ model in a transverse field has
been explored by using different criteria for detecting the entanglement \cite{Giampaolo2013,Hofmann2014}, which shows that it obeys
a scaling behavior near the critical point of
the quantum phase transition in the model.
Interest in the subject has increased since
the introduction of quantum discord \cite{Ollivier2001},
which is one of the most important quantum correlations that
characterizes the quantum phase transition \cite{Maziero2010}.
Following quantum discord, many other measures
have been introduced to explore the correlations in the context
of the $XY$ model. The quantum phase transition is studied using local quantum
uncertainty and Wigner-Yanase skew information
\cite{Cakmak2015}. It has been shown
that single-spin coherence reliably identifies
the quantum phase transition in
the thermal ground state of the anisotropic
spin-1/2 $XY$ chain in the transverse
magnetic field. Geometric discord
is used to characterize the quantum phase transition for the $XY$ model \cite{Cheng2012}.
There are many followed results dedicated to the quantum
phase transition in other spin chain models \cite{Altintas2012},
such as $XXZ$, $XYT$,
Lipkin-Meshkov-Glick (LMG), $XY$ with
the Dzyaloshinskii-Moriya (DM) interaction
\cite{Liu2011}, both in
the thermodynamic limit and in few body cases.

In this article, we consider more general quantum correlations in the $XY$ model.
The one-way quantum deficit is one of the
popular measures that can characterize and
quantify the quantum correlations \cite{Oppenheim2002}.
Nevertheless, the one-way quantum deficit
has not been studied with regards to characterizing
the $XY$ spin-1/2 chain. Quantum
coherence based on $l_1$ norm and relative entropy measures
is also a basic and important method for characterizing quantum systems
\cite{Baumgratz2014}.
We capture the quantum
phase transition using these measures to study quantum phase
transitions for the $XY$ chain in a transverse field.

The article is organized  as follows.
In Section II, we recall the
basic notation and concepts of the one-way
quantum deficit and quantum coherence.  The spin-1/2
anisotropic $XY$ chain is introduced in Section III. The numerical results regarding the quantum phase
transition are presented in Section IV. Finally, we present our conclusions in Section V.

\section{One-way quantum deficit and
quantum coherence}
Let us first review the basic definitions of
the one-way quantum deficit and quantum coherence.

{\sf One-way quantum deficit}~
The one-way quantum deficit is defined
as the difference in the von Neumann entropy of a bipartite state, $\varrho_{AB}$,
before and after a measurement is performed,
without a loss of generality, on particle $A$ \cite{Streltsov2011},
\begin{equation}\label{deficit}
	\Delta=\min_{\Pi_A^i} S(\sum_i\Pi_A^i(\varrho_{AB}))-S(\varrho_{AB}),
\end{equation}
where $\Pi_A$ is the measurement on
subsystem $A$ and $S(\varrho_{AB})=-{\rm Tr}\varrho_{AB} \log\varrho_{AB}$ is the
von Neumann entropy. Throughout the article, $\log$ is in base 2. The minimum is taken over
all local measurements $\Pi_A.$

{\sf Quantum coherence}~
We consider the $l_1$ norm and
relative entropy of coherence measures in this article.
For a fixed basis set $\{|i\rangle\}$,
the set of incoherent states $\mathcal{I}$
is the set of quantum states with diagonal
density matrices, with respect to this
basis.
For an arbitrary quantum state
\begin{equation}
	\varrho=\sum_{i,j}\varrho_{i,j}|i\rangle\langle j|,
\end{equation}
the \emph{$l_1$ norm} coherence,
$\mathcal{C}_{l_1}(\varrho)$, of the state $\varrho$ is defined by
\begin{equation}
	\mathcal{C}_{l_1}(\varrho)=\sum_{ i\ne j}|\varrho_{i,j}|,
\end{equation}
which is the sum of the absolute values of all the non-diagonal entries for $\varrho$.

The relative entropy of the coherence is defined as
\begin{equation}
	\mathcal{C}_{rel}(\varrho)=S(\varrho_{\rm diag})-S(\varrho),
\end{equation}
where $S(\cdot)$ is the von Neumann entropy and
$\varrho_{\rm diag}$ denotes the
state obtained from $\varrho$ by deleting all the
off-diagonal elements.

\begin{figure*}[htb]
\centering
\includegraphics[width=17cm]{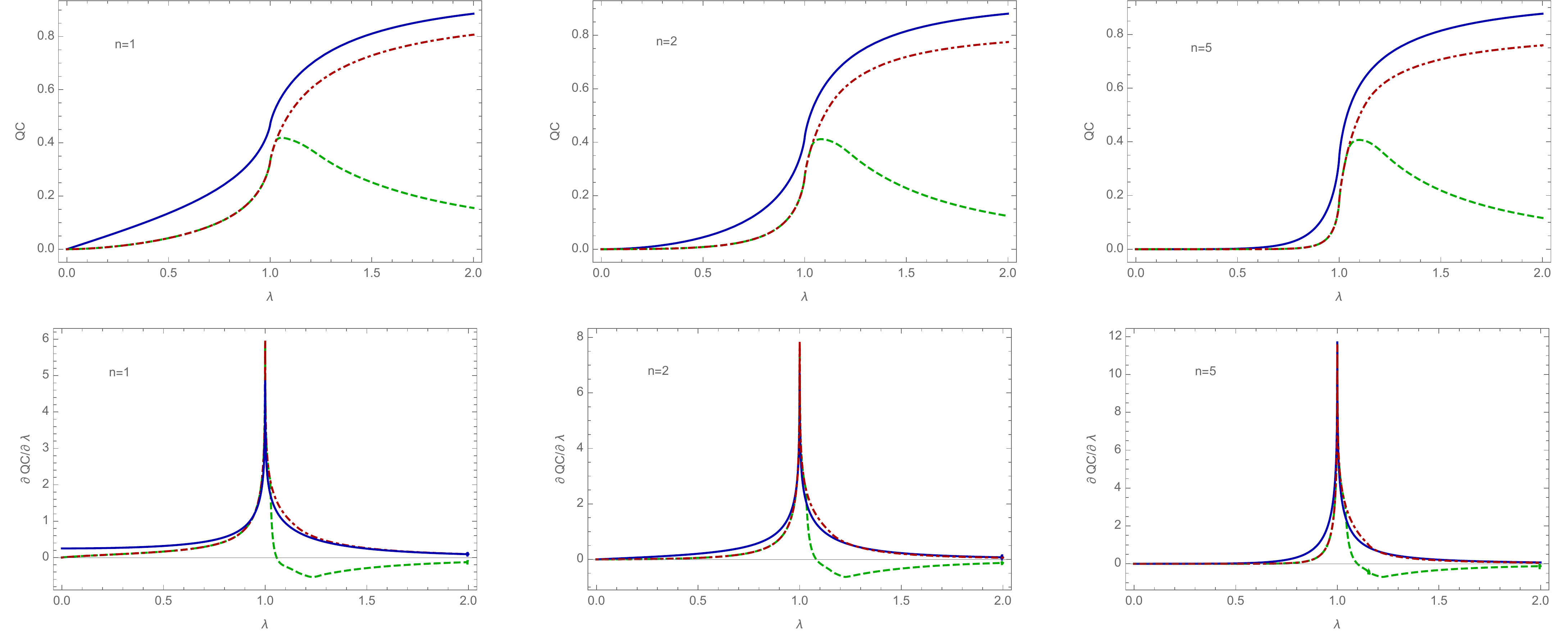}
\caption{{(Color online) Quantum correlation (QC)
behaviors of $XY$ chain at zero temperature.
Solid blue line denotes $l_1$
norm. Dotted-dashed red line describes
relative entropy of coherence. Dashed
green line depicts the one-way
quantum deficit, for the first, second and
fifth nearest neighbors ($n=1,2,5$ with $\gamma=0.5$) in the spin-1/2
chain in the transverse field, together with their first
derivations with respect to the parameter $\lambda$.}}
\label{nqpt}
\end{figure*}

\section{Anisotropic $XY$ chain}
The Hamiltonian, $H$, of the one-dimensional
anisotropic spin-$\frac12$ $XY$ chain
in a transverse magnetic field is given by
\begin{equation}
	H=-\sum_{j=0}^{N-1}\left\{
	\frac{\lambda}{2}[(1+\gamma)\sigma_j^x\sigma_{j+1}^x+(1-\gamma)\sigma_{j}^y\sigma_{j+1}^y]+
	\sigma_j^z\right\},
\end{equation}
where $\sigma_j^k (k=x,y,z)$ is the
$k$th component of the spin-1/2 Pauli operator
acting on the $j$th spin, $\gamma$
is the degree of anisotropy (for simplicity we
take this to be  $0\le\gamma\le1$),
and $\lambda$ is the strength of
the inverse of the external transverse
magnetic field. In this study, we focus on the infinite chain case, $N\rightarrow \infty.$
When $\gamma=0$, the Hamiltonian
reduces to the $XX$ chain, and the Ising model in transverse
field when $\gamma=1$.

The diagonalization procedure for the
$XY$ model includes the well-established
techniques of Jordan-Wigner and the Bogoliubov
transformation \cite{Barouch1970,Barouch1971}.
By considering the thermal ground state, the reduced density
operator for sites 0 and $n$ can be
described by
\begin{equation}\label{states}
	\rho_{0n}=\frac{1}{4}\left\{I_{0n}
	+\langle\sigma^z\rangle(\sigma_0^z
	+\sigma_n^z)+\sum_{k=x,y,z}
	\langle\sigma_0^k\sigma_n^k\rangle\sigma_0^k\sigma_n^k\right\},
\end{equation}
where $I_{0n}$ is the identity matrix
acting on the state space of sites
$0$ and $n$. Here, $n$ indicates for the distance between
two spins. The two-spin reduced
density matrix is only dependent on the
distance between the spins $n=|j-i|$,
with $j,i$ denoting two different spins.
The Hamiltonian exhibits
global $Z_2$ symmetry. The density
matrix is of the alphabet $X$ form,
\begin{widetext}
\begin{eqnarray}
\varrho_{0n}=\frac14
\left(\begin{array}{cccc}
1+2\langle\sigma^z\rangle+\langle\sigma_0^z\sigma_n^z\rangle & 0 & 0 & \langle\sigma_0^x\sigma_n^x\rangle-\langle\sigma_0^y\sigma_n^y\rangle \cr
0 & 1-\langle\sigma_0^z\sigma_n^z\rangle
& \langle\sigma_0^x\sigma_n^x\rangle+\langle\sigma_0^y\sigma_n^y\rangle	& 0 \cr
0 & \langle\sigma_0^x\sigma_n^x\rangle+\langle\sigma_0^y\sigma_n^y\rangle & 1-\langle\sigma_0^z\sigma_n^z\rangle & 0 \cr
\langle\sigma_0^x\sigma_n^x\rangle-\langle\sigma_0^y\sigma_n^y\rangle & 0 & 0 & 1-2\langle\sigma^z\rangle+\langle\sigma_0^z\sigma_n^z\rangle
\end{array}
\right). 	
\end{eqnarray}
\end{widetext}
Owing to the fact that the system is invariant under translations,
the entries of the two-site reduced-density
depend only on the distance, $n$. The transverse magnetization
is given by
\begin{eqnarray}\label{tm}
	\langle\sigma^z\rangle=-\int_0^\pi
	\frac{(1+\lambda\cos\phi)\tanh(\beta\omega_\phi)}{2\pi \omega_\phi}d\phi,
\end{eqnarray}
where $\omega_\phi=\sqrt{(\gamma \lambda\sin\phi)^2+(1+\lambda\cos\phi)^2}/2$ and $\beta=1/(kT)$
with $k$ being the Boltzmann's constant and $T$
the absolute temperature.
The two-point correlation function is given by
\begin{eqnarray}
	\langle\sigma_0^x\sigma_n^x\rangle=\left|
	\begin{array}{cccc}
	F_{-1} & F_{-2}& \cdots & F_{-n}\cr
	F_{0} & F_{-1}& \cdots & F_{-n+1}\cr
	\vdots & \vdots &\ddots & \vdots\cr
	F_{n-2} & F_{n-3}& \cdots & F_{-1}\cr		
	\end{array}
	\right|,
\end{eqnarray}
\begin{eqnarray}
	\langle\sigma_0^y\sigma_n^y\rangle=\left|
	\begin{array}{cccc}
	F_{1} & F_{0}& \cdots & F_{-n+2}\cr
	F_{2} & F_{1}& \cdots & F_{-n+3}\cr
	\vdots & \vdots &\ddots & \vdots\cr
	F_{n} & F_{n-1}& \cdots & F_{1}\cr		
	\end{array}
	\right|,
\end{eqnarray}
and
\begin{equation}
\langle\sigma_0^z\sigma_n^z\rangle=\langle\sigma^z\rangle^2-F_nF_{-n},	
\end{equation}
with
\begin{eqnarray}
F_n=&&\int_0^\pi\frac{\tanh(\beta\omega_\phi)}{2\pi\omega_\phi}\{\cos(n\phi)(1+\lambda\cos\phi)\nonumber\\
&&-\gamma\lambda\sin(n\phi)\sin\phi\}	d\phi.
\end{eqnarray}
Tracing out one of the two spins, we have
the reduced density matrix of a single-spin,
\begin{eqnarray}\label{ssrd}
	\varrho_0=\varrho_i=\frac12
	\left(\begin{array}{cc}1+\langle\sigma^z\rangle & 0\cr
	0 & 1-\langle\sigma^z\rangle
	\end{array}\right),
\end{eqnarray}
where $\langle\sigma^z\rangle$ is the
transverse magnetization  in Eq.(\ref{tm}). All of the
single-spin reduced density matrices
are of the same form (\ref{ssrd}).

\begin{figure*}[htbp!]
\centering
  \begin{tabular}{@{}cccc@{}}
    \includegraphics[width=.27\textwidth]{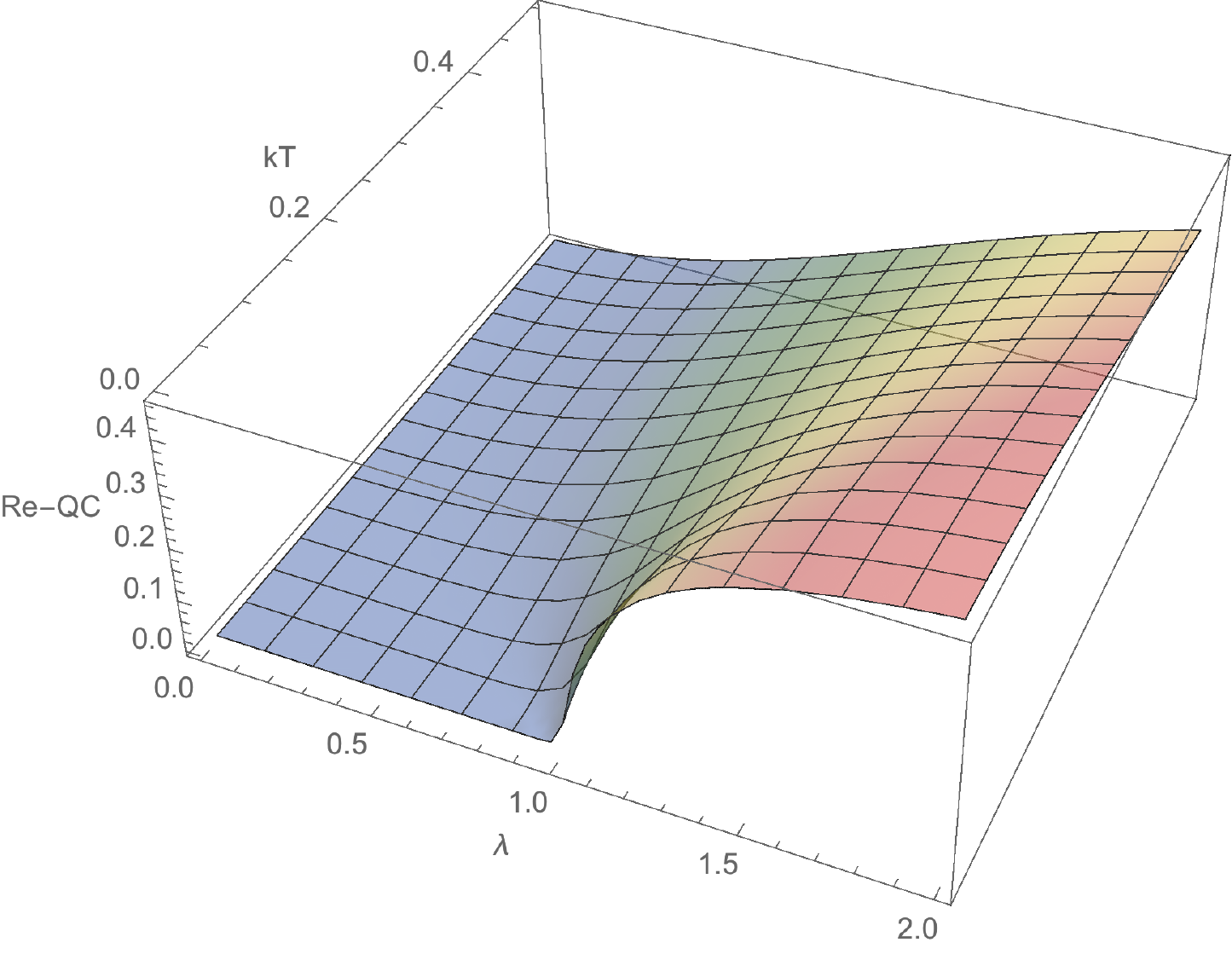} &
    \includegraphics[width=.27\textwidth]{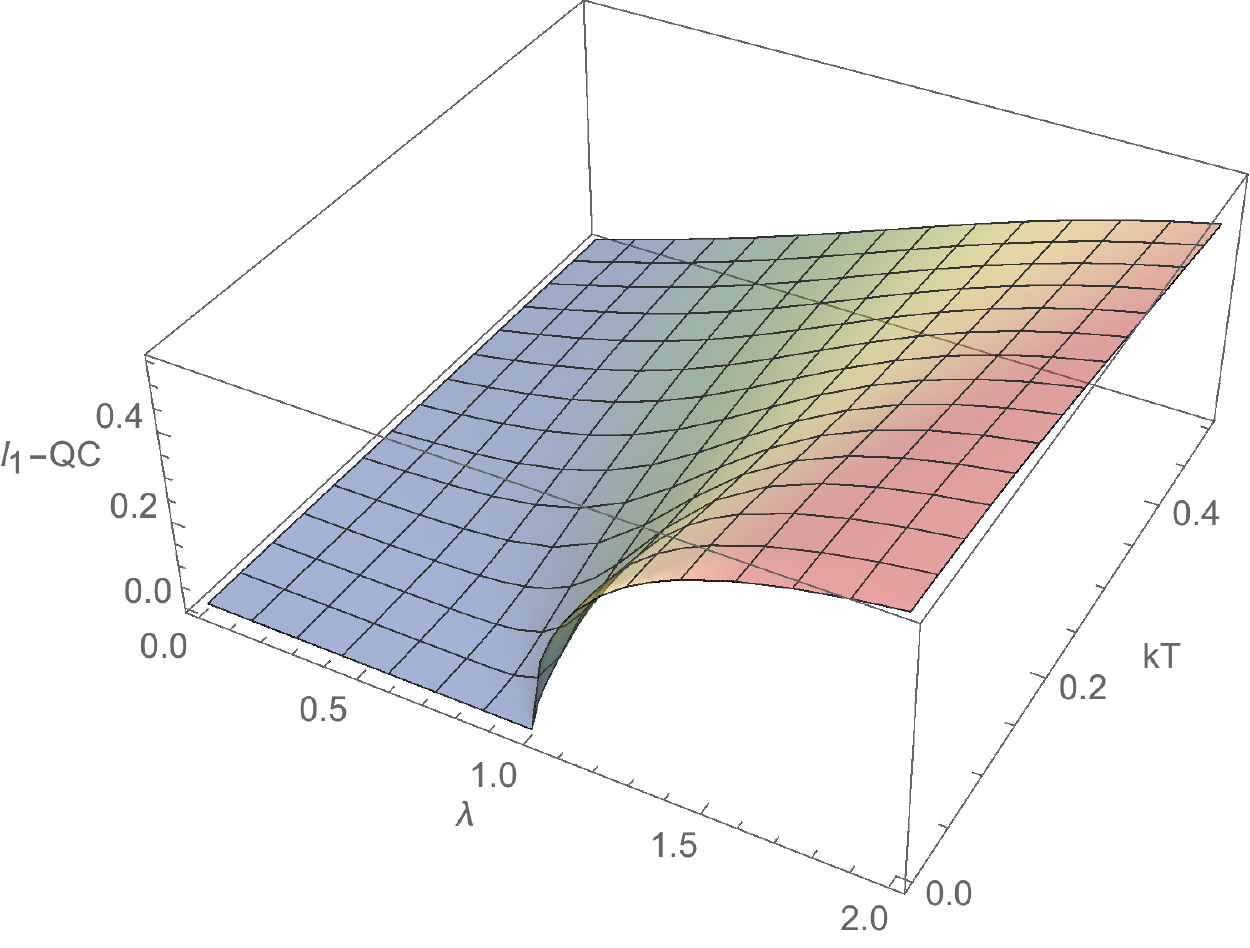} &
    \includegraphics[width=.27\textwidth]{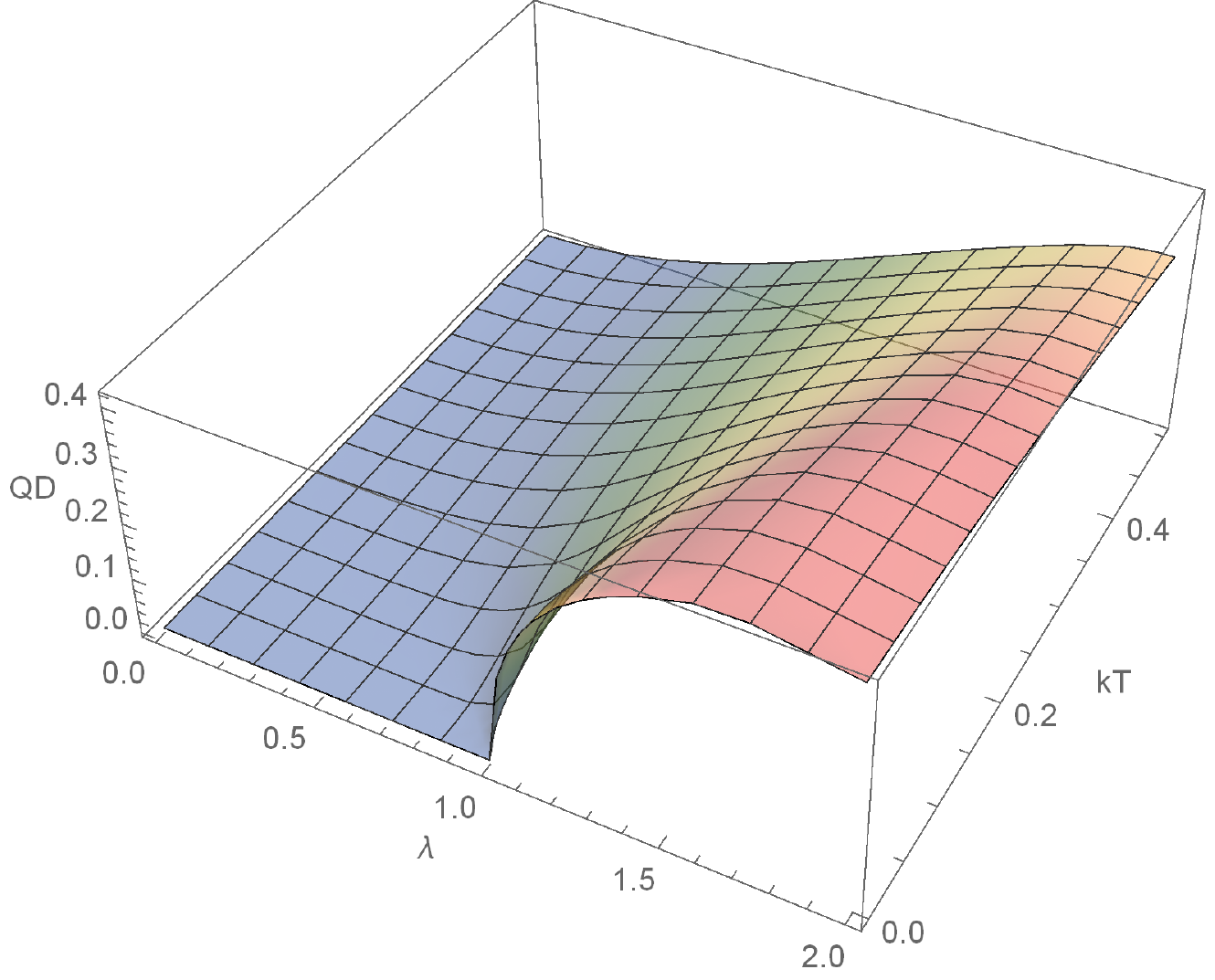}   \\
    \includegraphics[width=.27\textwidth]{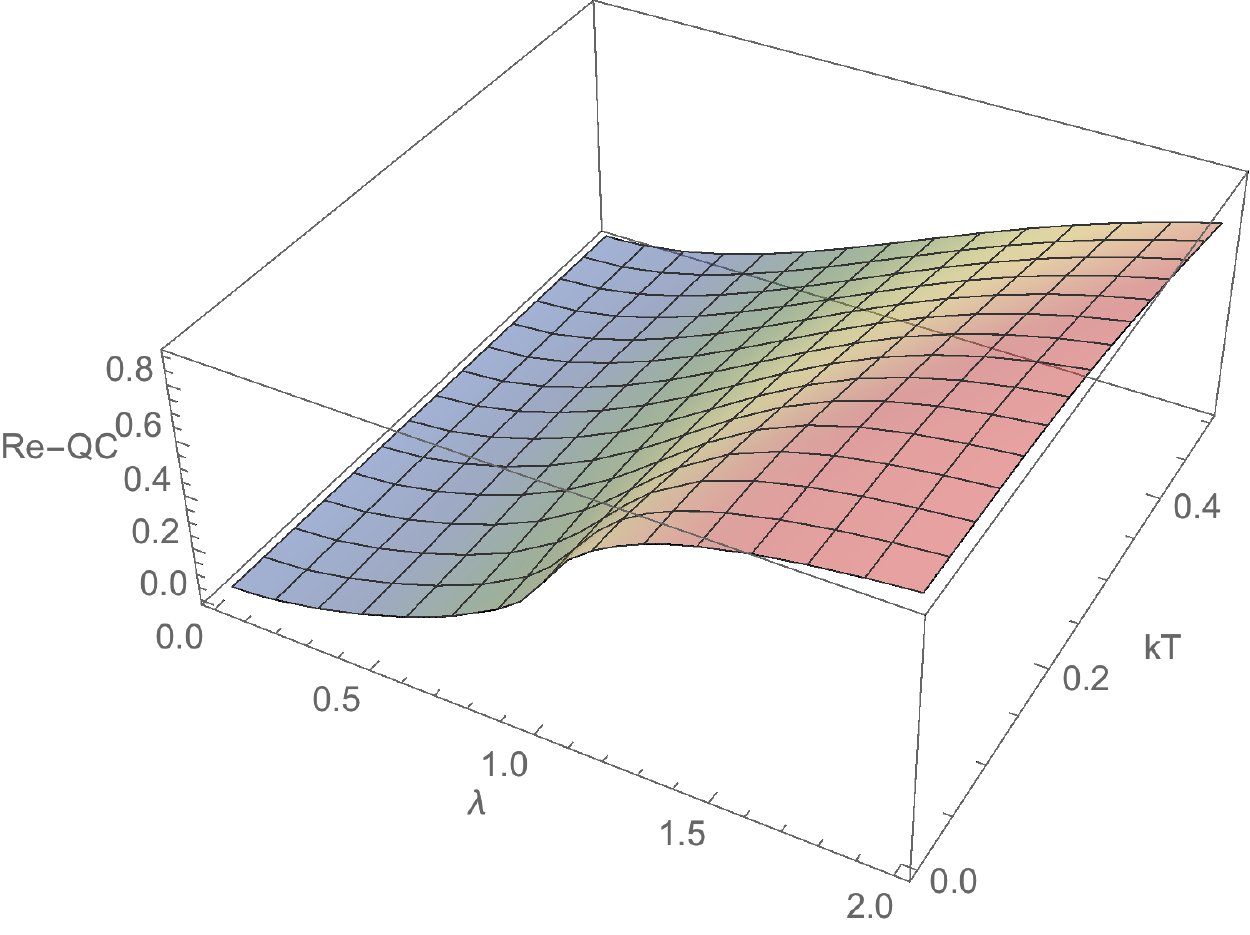} &
    \includegraphics[width=.27\textwidth]{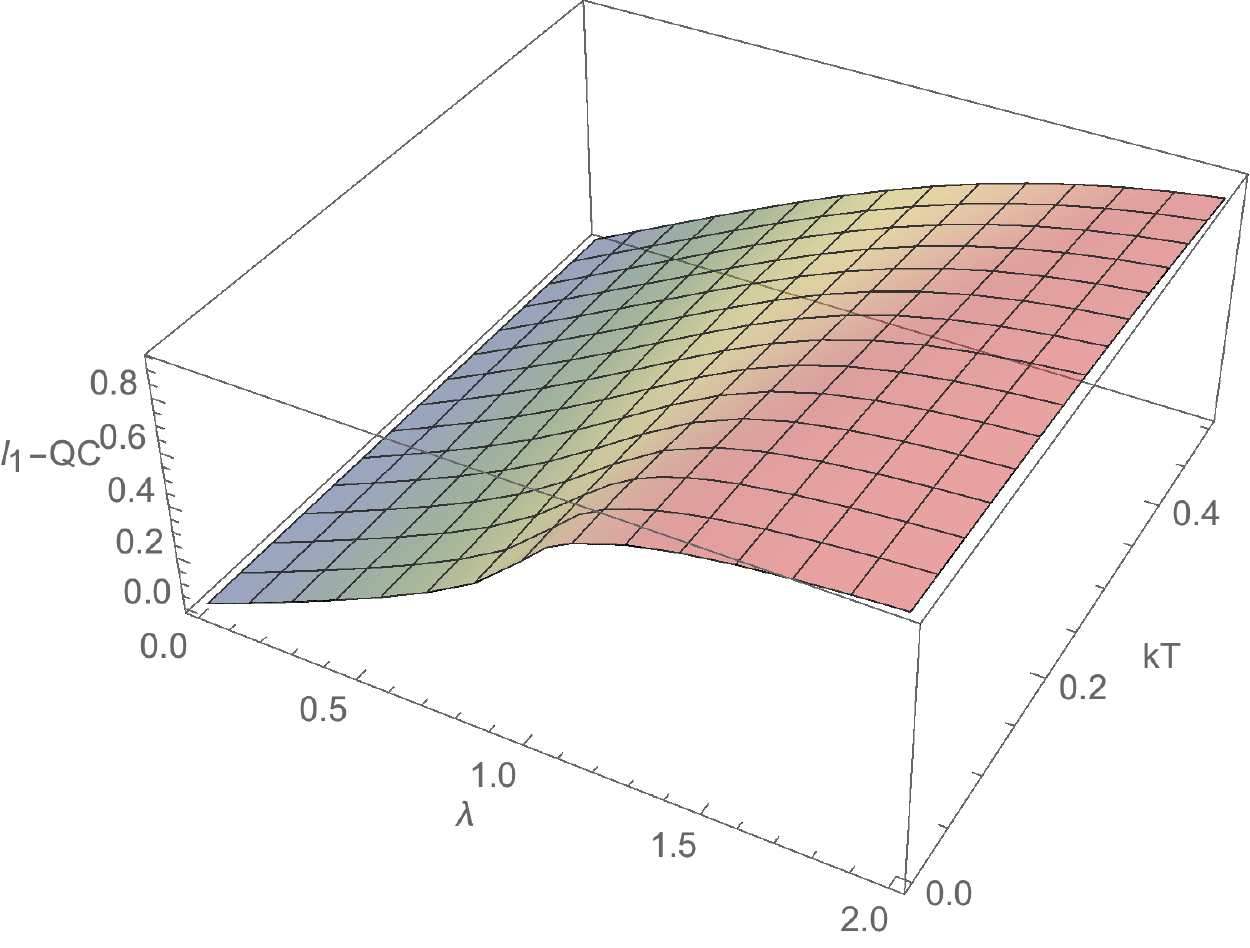} &
    \includegraphics[width=.27\textwidth]{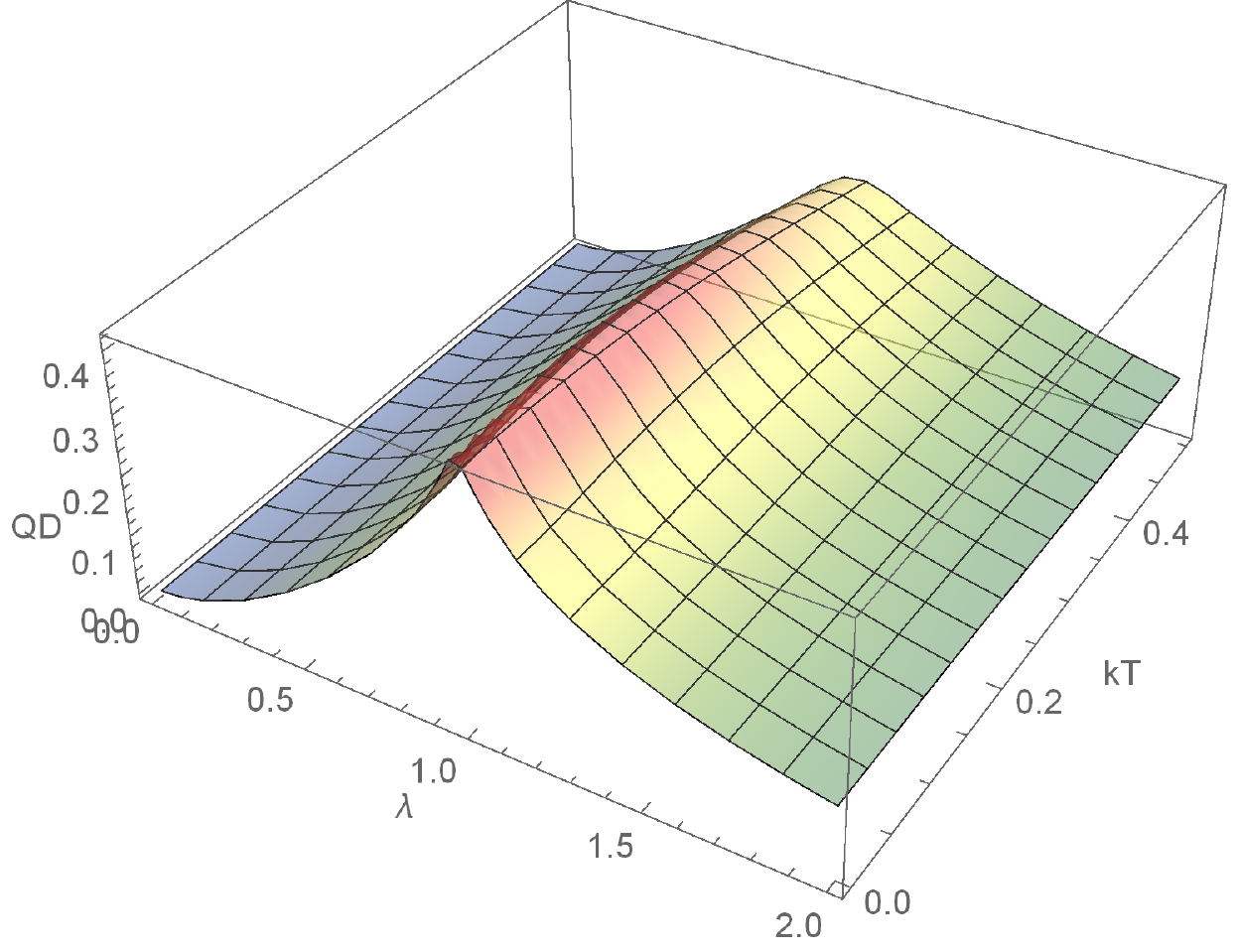}
  \end{tabular}
  \caption{(Color online) The first (second) row shows the quantum
  correlations (QCs) of the first neighbors spins for $XX$  (Ising) model.
  From left to right are the figures for
  relative entropy (Re-QC), $l_1$ norm ($l_1$-QC) and the
  one-way quantum deficit (QD), with respect to
  temperature ($kT$) and the magnetic
  field $\lambda$.}
  \label{finteT}
\end{figure*}

\section{Behaviors of correlations}
From the one-way quantum deficit defined in Eq.(\ref{deficit}), we have
\begin{equation}
S(\varrho_{0n})=-\sum_{i=0}^1(\eta_i\log\eta_i+\xi_i\log\xi_i),
\end{equation}
where
\begin{equation}
 	\eta_i=[1+\langle\sigma_0^z\sigma_n^z\rangle
 	+(-1)^i\sqrt{4\langle\sigma^z\rangle^2+(\langle\sigma_0^x\sigma_n^x\rangle-
 	\langle\sigma_0^y\sigma_n^y\rangle)^2}]/4,
\end{equation}
and
\begin{equation}
	\xi_i=[1-\langle\sigma_0^z\sigma_n^z\rangle
	+(-1)^i(\langle\sigma_0^x\sigma_n^x\rangle+\langle\sigma_0^y\sigma_n^y\rangle)]/4.
\end{equation}

We perform the complete set of orthonormal
projective measurements, $\Pi_n$, on the $n$th nearest spins,
with $\Pi_n^i=V|i\rangle\langle i|V^\dagger$,
where $i\in\{0,1\}$ and
\begin{eqnarray}
	V=\left(\begin{array}{cc}
		\cos(\theta/2) & e^{i\varphi}\sin(\theta/2)\cr
		-e^{-i\varphi}\sin(\theta/2) & \cos(\theta/2)
	\end{array}\right).
\end{eqnarray}
Thus, we obtain the first term of
Eq.(\ref{deficit}) as follows
\begin{eqnarray}
	S(\sum_i
\Pi_n^i(\varrho_{0n}))=-\sum_{i,j=0}^1\xi_{ij}\log\xi_{ij},
\end{eqnarray}
where
\begin{widetext}
\begin{eqnarray}
	\xi_{ij}=[1+(-1)^i\langle\sigma^z\rangle
	\cos\theta
	+(-1)^j\sqrt{[\langle\sigma_0^x\sigma_n^x\rangle^2\cos^2\varphi
	+\langle\sigma_0^y\sigma_n^y\rangle^2\sin^2\varphi]\sin^2\theta
	+[\langle\sigma^z\rangle+(-1)^i\langle
	\sigma_0^z\sigma_n^z\rangle\cos\theta]^2}]/4.
\end{eqnarray}	
\end{widetext}
Therefore, the one-way quantum deficit
is given by
$\{\theta, \varphi\}$
\begin{eqnarray}
	\Delta=\min_{\theta,\varphi}[-\sum_{i,j=0}^1\xi_{ij}\log\xi_{ij}]
	+\sum_{i=0}^1(\eta_i\log\eta_i+\xi_i\log\xi_i).\nonumber\\
\end{eqnarray}

The $l_1$ norm of the coherence can be directly shown to be
\begin{eqnarray}
	\mathcal{C}_{l_1}(\varrho_{0n})=\langle
	\sigma_0^x\sigma_n^x\rangle.
\end{eqnarray}
The relative entropy of the coherence is given by
\begin{equation}
	\mathcal{C}_{rel}(\varrho_{0n})=\sum_{i=0}^1 (-\zeta_i\log
	\zeta_i+\eta_i\log\eta_i+\xi_i\log\xi_i)-2\varepsilon\log\varepsilon,
\end{equation}
with
\begin{eqnarray}
	\zeta_i=[1+\langle\sigma_0^z\sigma_n^z\rangle+(-1)^i2\langle\sigma^z\rangle]/4
\end{eqnarray}
and $\varepsilon=[1-\langle\sigma_0^z\sigma_n^z\rangle]/4.$

Remarkably, the numerical analysis implies that the extremization is achieved when $\lambda\le1$. The one-way
quantum deficit can be represented by an analytical expression
by choosing $\theta=\varphi=0$, which is in
coincident with the relative entropy
of the coherence $\Delta=\mathcal{C}_{rel}(\varrho_{0n})$.
In the region where
$\lambda>1$, it is only possible to obtain the numerical
solution for the one-way quantum deficit.

\subsection{Quantum phase transition and correlations at zero
temperature}
The one-way quantum deficit, relative
entropy of coherence and $l_1$ norm
for the first, second and fifth nearest
neighbors in the thermal ground state
(\ref{states}) near zero temperature
are depicted in Fig.\ref{nqpt}.
The figures in the first row show the quantum coherence by the $l_1$ norm,
the relative entropy of coherence and the one-way quantum
deficit for $n=1,2$ and $5$, respectively, from left to right, with $\gamma=0.5$.
The figures in the second row are the first order
derivations of the corresponding quantum correlations with respect to the parameter $\lambda$.

As expected, all these three types of quantum correlations
decrease as the distance, $n$, increases. Nonetheless, we see a
clear difference between the regions where
$\lambda<1$ and $\lambda>1$. It is evident that
the first order derivations of the corresponding quantum correlations
with respect to the parameter, $\lambda$, are singular
at the quantum phase transition point $\lambda=1$. Namely, these quantum
correlations can be used to characterize the quantum
phase transition in this quantum system.

Some properties are obvious:
the quantum coherence via the $l_1$ norm is greater
than both the relative entropy of coherence and
the one-way quantum deficit.
The one-way quantum deficit coincides
with the relative entropy of
coherence in the region where $\lambda\le1$. While in the
region where $\lambda>1$, the $l_1$ norm
is greater than the relative entropy
of coherence, which is greater than
one-way quantum deficit.
A critical point appears at $\lambda=1$.
The first order derivations of these quantum correlations show
the quantum phase transition clearly: they
change sharply at $\lambda=1$ for $n=1,2$ and $5$, with $\gamma=0.5$.

\subsection{Quantum phase transition
and correlations at finite temperature}
We now consider the thermal state of the $XX$
($\gamma=0$) chain at finite temperatures.
In this case, we plot the three kinds of
quantum correlations for the first nearest neighbors, as shown in Fig.\ref{finteT}.
Similar results can be obtained
for the other nearest neighbors, $n$.
It can be seen that for a given
$\lambda$ $(\lambda>1)$, the quantum correlations decrease as the temperature increases. However, in a given region of $\lambda$ $(\lambda\le1)$,
the quantum correlations indeed increase as the temperature increases when near the critical value of $\lambda=1$. The quantum phase transition phenomenon disappears as
the temperature increases.

For the case of the transverse
Ising model ($\gamma=1$) at finite
temperatures, as can be seen in the figures in the second row of
Fig.\ref{finteT}, the one-way
quantum deficit increases when $\lambda$ increases until it is nearly 1. It
then decreases as $\lambda$ increases, even in the high temperature region. Meanwhile
the coherence always increases when $\lambda$ increases for a given temperature.

\section{Conclusions}
In this study, we investigated the
pairwise quantum correlations in the thermodynamic
limit of the anisotropic $XY$ spin-1/2
chain in the presence of an external
transverse magnetic field. The cases of both  zero
and finite temperatures have been considered. We have shown that the one-way
quantum deficit, the $l_1$ norm, and the
relative entropy of coherence can be used to
characterize the quantum phase transition
for the anisotropic $XY$ chain.
It has been shown that in the region where $\lambda\le1$, the one-way
quantum deficit has an analytical
expression that coincides with
the relative entropy of coherence.
The critical point is at $\lambda=1.$
The $XX$
chain and the Ising models have also been  studied at finite temperature.
These results can highlight the investigation of the relations between
quantum correlations and quantum phase transition.

\section{Acknowledgments}
This work is supported by the NSFC
under grant numbers 11275131, 11305105, 11675113 and Jiangxi Education Department Fund (KJLD14088).


\end{document}